\documentclass[sigconf, authorversion=true]{acmart}
\usepackage[english]{babel}
\usepackage[utf8]{inputenc}
\usepackage[T1]{fontenc}

\usepackage{graphicx}

\usepackage{multirow}

\usepackage{tikz}
\usetikzlibrary{fadings,trees,positioning,arrows,shapes}
\usepackage{etoolbox}

\usepackage[nomessages]{fp}

\usepackage{enumitem}
\setlist[itemize]{leftmargin=1em}
\setlist[description]{leftmargin=1em}

\newcommand{\mainTitle}{An Empirical Study of Configuration Mismatches in Linux}

\usepackage{flushend}

\usepackage{hyperref} 
\hypersetup{
    pdfauthor={Sascha El-Sharkawy, Adam Krafczyk, and Klaus Schmid},
    pdftitle={\mainTitle},
    pdfsubject={21st International Conference on Software Product Line},
    pdfkeywords={Software product lines; variability modeling; configuration mismatches; Linux; Kconfig; static analysis; empirical software engineering},
    pdfpagemode={UseOutlines},
    pdfproducer={LaTeX},
    colorlinks=true,
    linkcolor=black,          
    citecolor=black,          
    filecolor=black,          
    urlcolor=black            
}

\newcommand{\RQLayerOne}[1]{\textbf{RQ#1}}

\newcommand{\RQRefBrackets}[1]{(#1)}
\newcommand{\RQRef}[1]{\RQRefBrackets{\ref{#1}}}

\newcommand{\ATH}{\texttt{ATH\-5K}}
\newcommand{\ATHTwentyFive}{\texttt{ATH\-25}}
\newcommand{\ATHK}{\ATH\texttt{\_PCI}}
\newcommand{\vendorQualcomm}{\texttt{NET\_VEN\-DOR\_QUAL\-COMM}}
\newcommand{\QCA}{\texttt{QCA\-7000}}
\newcommand{\SERIAL}{\texttt{SE\-RI\-AL\_8250\_MID}}

\newcommand{\DMA}{\texttt{HSU\_DMA}}
\newcommand{\DMAPCI}{\texttt{HSU\_DMA\_PCI}}

\newcommand{\XEN}{\texttt{XEN}}
\newcommand{\XENVPMU}{\texttt{XEN\_HA\-VE\_VPMU}}
\newcommand{\XENSYS}{\texttt{XEN\_SYS\_HY\-PER\-VI\-SOR}}
\newcommand{\ifDef}{\texttt{\#ifdef}}
\newcommand{\ifDefBlock}{\ifDef-block}
\newcommand{\make}{\texttt{MAKE}}

\definecolor{BlueUp}{RGB}{163,196,255}
\definecolor{BlueMid}{RGB}{191,213,255}
\definecolor{BlueDown}{RGB}{229,238,255}

\newcommand{\nIneffective}{42}
\newcommand{\nSubmenus}{21}
\newcommand{\nInvisibles}{11}
\newcommand{\nHWvariables}{6}
\FPeval{\nSmells}{clip(\nIneffective + \nSubmenus + \nInvisibles + \nHWvariables)}
\FPeval{\nCritical}{clip(\nIneffective + \nInvisibles)}
\newcommand{\nVars}{68}
\newcommand{\nBools}{54}
\newcommand{\nTristates}{14}

\newcommand{\pieHigh}{4mm}

\newcommand{\slice}[6]{
  \pgfmathparse{0.5*#1+0.5*#2}
  \let\midangle\pgfmathresult

  \draw[very thin,fill=#4!#5!#6] (0,0) -- (#1:1) arc (#1:#2:1) -- cycle;
  \ifboolexpr{ test {\ifnumcomp{#1}{>}{50}}}
    {\shadedraw[bottom color=#4!#5!#6,top color=#4!5!black,draw=black,very thin]   
    (#1:1cm) --++(0,-\pieHigh) arc (#1:#2 :1cm) -- ++(0,\pieHigh)  arc (#2 :#1  :1cm) -- cycle; }
    {\shadedraw[bottom color=#4!#5!#6,top color=#4!5!black,draw=black,very thin]   
    (50/100*360:1cm) --++(0,-\pieHigh) arc (50/100*360:#2 :1cm) -- ++(0,\pieHigh)  arc (#2 :50/100*360 :1cm) -- cycle; }

  \pgfmathparse{min((#2-#1-10)/110*(-0.3),0)}
  \let\temp\pgfmathresult
  \pgfmathparse{max(\temp,-0.5) + 0.8}
  \let\innerpos\pgfmathresult
  \node at (\midangle:\innerpos) {#3};
}

\newcommand{\pieLabel}[4]{
  \coordinate (point) at (#1);
  \node[#4](label) at (#2){#3};
  \draw[dashed] (point) -- (label);
  \fill(point) circle(.3mm);
}

\newcommand{\smellDesc}[6]{node (#1)[#2] {#3\\#4\\C:\ #5,\hfill Kb:\ #6}}
\newcommand{\categoryDesc}[6]{node (#1)[#2] {#3\\#4}}

\usepackage{listings}
\lstdefinelanguage{config} {
  language=C,
  morecomment=[l]{\#},
  morekeywords=[2]{y, n, m},
  keywordstyle=[2]{\color{red!70!black}}
}

\lstdefinelanguage{kconfig} {
  language=config,
  string=[b]",
  keywords = {choice, endchoice, config, bool, tristate, string, int, hex, select, depends, on, option, modules, default, def_bool, def_tristate, range, if, endif, menu, endmenu, prompt, source, help}
}

\lstdefinelanguage{kbuild} {
  language=[gnu]make,
}

\lstdefinelanguage{cWithPre} {
  language=C,
  morecomment=[l]{\#}
}

\lstdefinelanguage{pseudo} {
    language=Java,
    morekeywords=[2]{end, done, begin},
    mathescape
}
\usepackage{ifthen}
\newboolean{showToDos}
\setboolean{showToDos}{false} 
\ifshowToDos
  \usepackage{marginnote} 
  \let\marginpar\marginnote
  
  \newcommand{\ToDo}[1]{\noindent\fcolorbox{black}{yellow!20!white}{\parbox{.975\columnwidth}{#1}}}
  \newcommand{\SubjectedToDo}[2]{\ToDo{\textbf{#1:}#2}}
  \newcommand{\ToDoInline}[1]{\fcolorbox{black}{yellow!20!white}{#1}}
  \newcommand{\Outline}[1]{\color{blue}{#1}\color{black}}
  \newcommand{\Problem}[1]{\color{red}{#1}\color{black}}
  \newcommand{\secSize}[1]{[#1]}
  
  \usepackage[textsize=small,colorinlistoftodos]{todonotes}
  \newcommand{\ks}[1]{\todo[color=orange!20,linecolor=orange!60,bordercolor=orange!60]{\textbf{KS:} #1}}
  \newcommand{\se}[1]{\todo[color=red!20,linecolor=red!60,bordercolor=red!60]{\textbf{SE:} #1}}
  \newcommand{\ak}[1]{\todo[color=blue!20,linecolor=blue!60,bordercolor=blue!60]{\textbf{AK:} #1}}
  
  \usepackage{pifont}
  
\else
  \newcommand{\ToDo}[1]{}
  \newcommand{\SubjectedToDo}[2]{}
  \newcommand{\ToDoInline}[1]{#1} 
  \newcommand{\Outline}[1]{}
  \newcommand{\Problem}[1]{}
  \newcommand{\secSize}[1]{}
  
  \newcommand{\ks}[1]{}
  \newcommand{\se}[1]{}
  \newcommand{\ak}[1]{}
\fi

\begin{document}
\title{\mainTitle}

\author{Sascha El-Sharkawy}
\affiliation{
  \institution{University of Hildesheim,\\Institute of Computer Science}
  \streetaddress{Universitätsplatz 1}
  \city{31141 Hildesheim} 
  \state{Germany} 
}
\email{elscha@sse.uni-hildesheim.de}

\author{Adam Krafczyk}
\affiliation{
  \institution{University of Hildesheim,\\Institute of Computer Science}
  \streetaddress{Universitätsplatz 1}
  \city{31141 Hildesheim} 
  \state{Germany} 
}
\email{adam@sse.uni-hildesheim.de}

\author{Klaus Schmid}
\affiliation{
  \institution{University of Hildesheim,\\Institute of Computer Science}
  \streetaddress{Universitätsplatz 1}
  \city{31141 Hildesheim} 
  \state{Germany} 
}
\email{schmid@sse.uni-hildesheim.de}



\begin{abstract}
Ideally the variability of a product line is represented completely and correctly by its   variability model.
However, in practice additional variability is often represented on the level of the build system or in the code. 
Such a situation may lead to inconsistencies, where the actually realized variability does not fully correspond to the one described by the variability model. 
In this paper we focus on configuration mismatches, i.e., cases where the effective variability differs from the variability as it is represented by the variability model. 
While previous research has already shown that these situations still exist even today in well-analyzed product lines like Linux, so far it was unclear under what circumstances such issues occur in reality. In particular, it is open what types of configuration mismatches occur and how severe they are. 
Here, our contribution is to close this gap by presenting a detailed manual analysis of 80 configuration mismatches in the Linux 4.4.1 kernel and assess their criticality. We identify various categories of configuration issues and show that about two-thirds of the configuration mismatches may actually lead to kernel misconfigurations.
\end{abstract}

%
%





\setcopyright{acmlicensed}

\acmDOI{http://dx.doi.org/10.1145/3106195.3106208}

\acmISBN{978-1-4503-5221-5/17/09}

\acmConference{SPLC '17}{September 25--29, 2017}{Sevilla, Spain}
\acmYear{2017}
\copyrightyear{2017}

\acmPrice{15.00}

\maketitle

\vspace{1em}
\section{Introduction}
In software product line engineering, dependency management, which is part of variability modeling, is not trivial to get right, but fundamental to identify the valid configurations. The complexity of defining the variability and their dependencies to the implementation consistently can be illustrated by the fact that, for instance, Linux version 4.4.1 has more than 15,500 configurable variables to control the conditional inclusion of over 40,000 source files with approximate 102,000 preprocessor-dependent code blocks \cite{CMDetector}. Thus, the variability model must be carefully maintained to reflect the implementation as otherwise the configuration process yields defective products. In recent years, much effort was spent in research to detect conflicts between the variability model and implementation artifacts. 
In this paper, we use the technique of \textit{feature effect}s \cite{NadiBergerKastner+15} to extract constraints from code and build artifacts and check if they are covered by the variability model. If the variability model does not cover these feature effect constraints, the configuration process becomes further restricted through these feature effects in an invisible manner. We call this divergence between modeled and implemented dependencies, \textit{configuration mismatches} \cite {El-SharkawyKrafczykSchmid16}. We present an empirical study of those divergences to better understand their sources and to assess their criticality. While the basic identification of mismatches was automated, the assessment was done manually as it required the interpretation of the situation.


The identified mismatches point to situations where the implementation of the product line introduces additional dependencies not covered by the variability model. In Linux, these dependencies do not necessarily lead to compilation errors but bear the risk of deriving the wrong product without any notification. Similar problems can be expected for other product lines using preprocessor-based instantiation, a widespread technique in industrial product lines \cite{HunsenZhangSiegmund+16}. 
In this study we answer the following research questions:
\begin{enumerate}[label=\RQLayerOne{\arabic*}]
  \item \label{rq:severity} What is the severity of configuration mismatches?
  \item \label{rq:technical} What are the technical characteristics of variables, which cause configuration mismatches? 
  \item \label{rq:characteristics} How do configuration mismatches differ conceptually?
  \item \label{rq:spaces} Are all three spaces: variability model, code artifacts, and build space always involved in a configuration mismatch? If not, what is their relation?
\end{enumerate}

\vspace{-1ex}
While earlier studies presented technical approaches for identifying mismatches, they did not provide an understanding of their causes, as they never performed a manual analysis. Here, our contribution is to provide an empirical analysis contributing to the understanding of these configuration mismatches. As a result, we identified four main types of configuration mismatches. We discuss each type based on examples taken from our Linux-analysis and identify the severity of the different mismatches. The severity ranges from rather unproblematic mismatches, e.g., in cases where variabilities were only introduced for structuring purposes to cases with mismatches of high severity, where the resulting system may be functionally different from the expected one. We could also identify that 66\% of the mismatches must be regarded as severe.

The remainder of this paper is structured as follows. In the next section we report related work. Section~\ref{sec:context} introduces our object of investigation: Linux. We explain the technical background, which is necessary to understand our findings. In Section~\ref{sec:setup}, we introduce our experimental setup. In Section~\ref{sec:analysis}, we present the results of our manual analysis while we discuss our findings in Section~\ref{sec:Discussion}. Section~\ref{sec:threats} discusses threats to validity, before we conclude and outline future work in Section~\ref{sec:Conclusion}.

\vspace{-1ex}
\section{Related Work \secSize{1/2 - 3/4}}
\label{sec:Related Work}
In this paper, we compare the variability information of code assets with the information in the variability model to detect mismatches in the dependency management. We then analyze the findings to present potential reasons for these configuration mismatches. This is done based on the Linux kernel as an use case for a complex real world software product line. We identified two categories of papers, which are relevant to this work: General papers about \textit{defect analysis} of variability and papers \textit{analyzing the characteristics of Linux}. The remainder is organized according to these two categories.

\textbf{Defect analysis.} In recent years, much effort was spent to detect defects related to software product line development, like (un-)dead features \cite{BenavidesSeguraRuiz-Cort10}, (un-)dead code \cite{TartlerLohmannSincero+11}, configuration coverage \cite{TartlerDietrichSincero+14}, or variability-aware type checking and liveness analysis \cite{LiebigRheinKastner+13}. Also Thüm et al.~\cite{ThumApelKastner+14} surveyed general analysis strategies for software product lines. Here, we identify and analyze configuration mismatches contrary to the identification of permanent unconfigurable blocks or conditional parsing errors.

In \cite{El-SharkawyKrafczykSchmid16}, we presented already an efficient algorithm to identify a specific kind of configuration mismatches; missing nesting dependencies. In this paper, we use the technique of feature effects \cite{NadiBergerKastner+15, NadiBergerKastner+14} to allow also the detection of configuration mismatches in cross-tree constraints. While this paper focuses the manual inspection of configuration mismatches to identify real problems.

\textbf{Analyzing the characteristics of Linux.} The kernel maintainers developed their own solutions for variability management (\textit{Kconfig}) and instantiation (\textit{Kbuild}), which we introduce in the next section. These two techniques have been analyzed in many scientific studies. In Section~\ref{sec:setup}, we present our approach for detecting configuration mismatches, which reuses the findings of the work presented here.

\textit{Kconfig} was developed as a textual variability language to manage the variability of Linux \cite{KConfig_Language}. While textual variability languages from a scientific background have already been extensively compared \cite{EichelbergerSchmid14c}, we currently only know of a comparison of eCos and Kconfig \cite{BergerSheLotufo+13}. Other work analyzes the capabilities of Kconfig in isolation. A prerequisite for many analysis is the translation of Kconfig models to a logical representation like propositional logic.  This translation is not a trivial task, since the semantics are rather unclear because of many corner cases. In~\cite{El-SharkawyKrafczykSchmid15, El-SharkawyKrafczykAsad+15} we analyzed the quality of tools regarding the translation of Kconfig models. These results formed the basis for the selection of an appropriate translation tool in Section~\ref{sec:toolchain}.

\begin{figure}[tb]
	\centering
\begin{lstlisting}[language=kconfig,caption={Example of a Kconfig file (indentation not necessary, only used for better readability).},label=lst:kconfig example]
config BOOLEAN_VAR1
    bool "Visible Boolean"
    select BOOLEAN_VAR2
    
config TRISTATE_VAR
    tristate "Visible Tristate" 
    depends on BOOLEAN_VAR
    default m
            
config BOOLEAN_VAR2
    bool
    default y if TRISTATE_VAR
\end{lstlisting}
  \vspace{-1em}
\end{figure}

\textit{Kbuild} \cite{KBuild_make} is the build process of Linux and responsible for the mapping of the configuration to the code files. It makes use of more than 70 \% of all Kconfig variables \cite{CMDetector}. This makes the analysis of Kbuild an important part for any comprehensive analysis of the variability of Linux. Two strategies exist for the analysis of Kbuild, which is in principle turing-complete since it is realized with \make. Dietrich et al.\ \cite{DietrichTartlerSchroderPreikschat+12} query the build system to probe which files are controlled by which variables of the variability model. They state that this concept is very robust as it works without manual intervention for all architectures on all versions of Linux, but needs approximately 90 minutes to analyze one architecture due the amount of probing steps. Nadi et al.\ \cite{NadiHolt11, NadiHolt12} as well as Berger et al.\ \cite{BergerSheLotufo+10, KBuildMiner} use text-based analysis of all build scripts. A comparison of all three implementations show that the the tools by Dietrich et al.\ and Berger et al.\ have the highest accuracy \cite{DietrichTartlerSchroderPreikschat+12}.



\section{Context \secSize{1}}
\label{sec:context}
Configurable systems, like Linux, are usually divided into \textit{problem space} and \textit{solution space} \cite{CzarneckiEisenecker00}. The problem space describes the range of supported products without any implementation details, i.e., the variability model. The solution space comprises the artifacts and their instantiation rules. Below, we briefly introduce the relevant parts of Linux related to the remainder of this paper. 

\subsection{Problem Space}
In Linux the problem space is modeled with Kconfig \cite{KConfig_Language, El-SharkawyKrafczykSchmid15}. Kconfig is a textual language, which was developed specifically for Linux and managing its variability. Although it never became a standalone project, it was also used in several other projects \cite{BergerSheLotufo+12}. We describe only the most important variability management concepts relevant to explain our findings in Section \ref{sec:analysis}. This is done based on the example given in Listing~\ref{lst:kconfig example}.

The Linux kernel supports multiple CPU architectures. Version 4.4.1 of the kernel supports already 31 different architectures. Each of these architectures has its own Kconfig files, which serve as start point. These architecture-specific models specify architecture-specific variables, settings, and constraints. Most of the Kconfig model is architecture independent and included by the architecture-specific start files. However, the architecture-specific start files require that each architecture is analyzed in isolation.\se{Evtl. noch choices, da es in Section~\ref{sec:analysis} als ein Fehler unserer Analyse angesprochen wird.}


\textit{Config options} are the main entities of the language and can be seen as variables. These config options can be of type \texttt{tristate}, \texttt{bool}, \texttt{string}, \texttt{hex}, or \texttt{int}. All but the first are known from typical programming languages and have the same semantics. Tristate config options encode the following alternatives: \texttt{n} (the related feature will not be part of the resulting system), \texttt{y} (the related feature will be a permanent part of the resulting system), and \texttt{m} (the related feature will be compiled as a module, which means that it can be flexibly loaded or unloaded at runtime). An analysis of Linux has shown that most config options are either of type \texttt{bool} (36\%) or \texttt{tristate} (58\%) \cite{BergerSheLotufo+13}. For this reason, we focus only on these two types as they are also sufficient to understand the findings of Section~\ref{sec:analysis}. Lines~\ToDoInline{5--8} of Listing~\ref{lst:kconfig example} declare a tristate config option.

Config options can be augmented with \textit{(conditional) attributes}, which are described below:

A \textit{prompt} displays a name to the user. A config option without a prompt is invisible and cannot be altered directly (only via constraints). For instance, prompts can be defined together with the type definition of the config option (Lines~\ToDoInline{2} and \ToDoInline{6}).

The \textit{default} attribute is used to specify default values of config options (Line~\ToDoInline{8}). These defaults can be changed by the user if the config option is visible. A config option can contain any number of conditional defaults (Line~\ToDoInline{12}).

Kconfig specifies two different kinds of \textit{constraints} to restrict the possible values of a config option: \texttt{depends on} and \texttt{select}. These constraints are also modeled as attributes of config options:

\texttt{depends on} constraints are used to describe whether a config option can be configured or is disabled. Contrary to the other attributes, this attribute can not be made conditional (Line~\ToDoInline{7}). 

\texttt{select} constraints are used to specify a lower bound (\texttt{y} > \texttt{m} > \texttt{n}). The current value of the surrounding config option is used as a minimal value for the selected config option. It is also possible to select a config option without fulfilling its \texttt{depends on} constraints. In this case, the \texttt{depends on} constraints are ignored. For this reason, the kernel developers discourage the selection of visible variables or variables with dependencies to avoid \textit{``illegal configurations''} \cite {KConfig_Language}. As most attributes, \texttt{select} constraints may be conditional. In such a case, the \texttt{select} constraint has only an effect if its condition is fulfilled. Line~\ToDoInline{3} models an unconditional \texttt{select} constraint; as long as \texttt{BOOLEAN\_VAR1} is selected \texttt{BOOLEAN\_VAR2} will also permanently be set to \texttt{y}.

\begin{figure*}[t]
	\centering
		\includegraphics[scale=0.565]{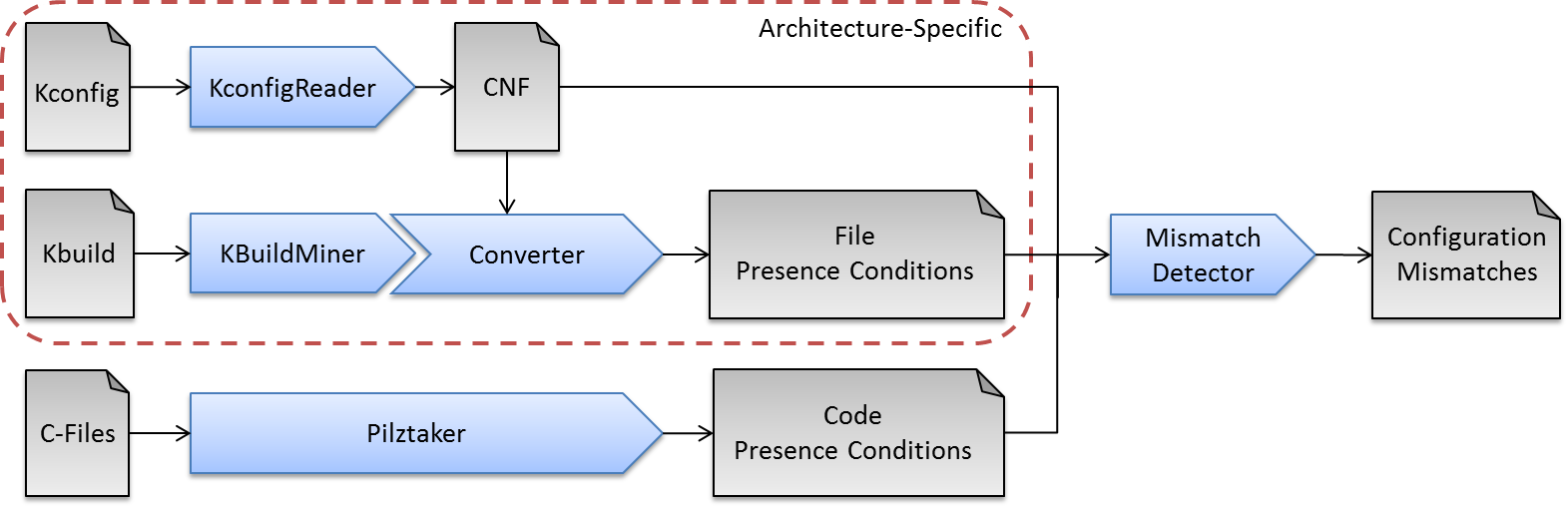}
  \vspace{-1ex}
  \caption{Our toolchain for the detection of configuration mismatches.}
  \label{fig:Architecture}
  \vspace{-1em}
\end{figure*}

\vspace{-1ex}
\subsection{Solution Space \secSize{2-3 Absätze}}
The Linux kernel is mostly implemented in C and some smaller parts in assembler \cite{KBuild_make}. Variability is realized through preprocessor statements. In our current analysis, we focus on C-code files only. 

The build process of the Linux kernel is realized through \textit{Kbuild}, which is implemented with \make. It is separated in a collection of Makefiles. Also these files contain variability; C-files may be excluded, linked statically into the final product, or be compiled as loadable modules depending on the configuration \cite{NadiHolt11}. Further, Kconfig variables may be used to include or exclude complete subfolders including their Kbuild scripts and implementation corresponding to Linux subsystems \cite{KBuild_make}.

All files of the solution space add the prefix \texttt{CONFIG\_} to refer to variables of the Kconfig model. This simplifies the differentiation between Kconfig variables and elements like \#include guards or compiler flags.

\section{Experimental Set-Up \secSize{1 - 1,5}}
\label{sec:setup}

This section describes our approach of identifying configuration mismatches. We define configuration mismatches and explain how they can be detected in a software product line, then we describe our toolchain for applying this concept to the Linux kernel.

\subsection{Conceptual Foundation}

Configuration mismatches point to a divergence between configuration information in the variability model and implemented variability of (code) assets \cite{El-SharkawyKrafczykSchmid16}. More precisely, this means that the solution space further restricts user-defined configurations through unmodeled dependencies. For the detection of configuration mismatches, we use the approach of \textit{feature effects} as described in \cite{NadiBergerKastner+15}, which expresses under which condition a variable has an impact on the product derivation. In their evaluation, Nadi et al.\ checked whether the identified feature effects are also contained in the variability model. They recognized that approximate 4\% of the identified feature effects are not part of the variability model, that is what we define as a configuration mismatch \cite{El-SharkawyKrafczykSchmid16}. However, they did not apply a manual analysis of the identified configuration mismatches. We reproduce their experimental set-up and apply a manual analysis of the identified configuration mismatches. In this section, we give an introduction to the concept of feature effect identification as defined by Nadi et al.\ \cite{NadiBergerKastner+15}.

A feature effect ($FE_{v}$) for a variable $v$ of the variability model is a combination of all presence conditions, which make use of $v$. Based on \cite{NadiBergerKastner+15}, we use the term \textit{presence condition (PC)} to refer to a propositional expression over Kconfig variables that determines when a certain code artifact is compiled. This comprises very fine-grained \ifDefBlock{s} surrounding a single line in a C-file, but also the definition of conditional compilation of complete C-files in Makefiles. For this reason, we use for the identification of feature effects the presence conditions from code files as well as from Kbuild.

For a variability variable $v$ and $PCs_{v}$ the presence conditions it is used in, the feature effect is defined as \cite{NadiBergerKastner+15}\footnote{$PC[v \leftarrow C]$ means replacing each occurrence of $v$ in $PC$ with constant $C$}:

\begin{equation}\label{eq:feature effect}
  FE_{v} \coloneqq \bigvee_{PC\ \in\ PCs_{v}} PC[v \leftarrow True] \oplus PC[v \leftarrow False]
\end{equation}

The approach of Formula~\ref{eq:feature effect} is based on propositional logic and, thus, does not support tristate logic directly. For this reason, tristate variables are translated into two variables and one constraint. The constraint is used to allow only the same configurations in the translated formulas as in the original variability model. The translation into 2 distinct variables makes it necessary to compute the feature effects for both translated variables separately.

Ideally, the variability model allows the configuration of $v$ only if it affects the final product. $v \rightarrow FE_{v}$ specifies under which condition ($FE_{v}$), the selection of $v$ has an effect on the final product. It can then be checked whether the variability model ($VM$) implies (i.e.\ contains) this constraint:

\begin{equation}\label{eq:tautology}
  \models \bigg( VM \rightarrow ( v \rightarrow FE_{v} ) \bigg)
\end{equation}

A configuration mismatch is detected, if this condition does not hold, that is if and only if the following formula is satisfiable:
\begin{equation}\label{eq:sat test}
  SAT( VM \land v \land \lnot FE_{v} )
\end{equation}

In this case, every valid solution (considering the variability model) of $\lnot FE_{v}$ is then a partial configuration in which selecting or deselecting $v$ has no effect on the final product. These partial configurations formed the basis for our manual analysis in Section~\ref{sec:analysis}, because they are useful for tracking down and deciding whether there is an underlying problem that needs to be fixed. The distinct calculation of feature effects for the permanent selection (\texttt{y}) and the selection as module (\texttt{m}) for tristate variables facilitates an independent analysis for both selections. However, it also doubles the number of reports in case that both assignments are problematic.

\subsection{Toolchain}
\label{sec:toolchain}
For our experiment, we analyzed the x86 architecture of the Linux kernel version 4.4.1. Figure~\ref{fig:Architecture} presents the toolchain, which we used for our analysis. For each location of variability (Kconfig, Kbuild, and C preprocessor), we used different, already available extraction tools. The outputs of the three tools are then combined and analyzed in a tool, that is specifically written for this task.

For the translation of the \textit{Kconfig} model to propositional logic, we chose KconfigReader \cite{KConfigReader}. This allows us to use SAT-solvers later on. In contrast to the other available tools for the same task, KconfigReader produces the most reliable results \cite{El-SharkawyKrafczykSchmid15}. For our analysis, we used the latest commit from 01.07.2016\footnote{commit hash \href{https://github.com/ckaestne/kconfigreader/tree/913bf3178af5a8ac8bedc5e8733561ed38280cf9}{\texttt{913bf31}}}.

For the extraction of presence conditions from \textit{Kbuild} we chose KbuildMiner \cite{KBuildMiner}. This tool computes for each file under which condition it is compiled, regarding a specific architecture. The result is then converted into a format, which is compatible with the other tools. This conversion also considers the information from Kconfig to differentiate between tristate and boolean variables. It is important to note, that KbuildMiner parses the Kbuild files only heuristically, so it is not completely reliable. However, a comparison shows that KbuildMiner is as reliable as the other available tools \cite{DietrichTartlerSchroderPreikschat+12}. We opted for using KbuildMiner since we had good experiences with this tool in the past and since it has a significantly lower runtime than Golem (under a minute compared to about 90 minutes) while having comparable reliability. For our analysis, we used the latest commit from 01.07.2016\footnote{commit hash \href{https://github.com/ckaestne/KBuildMiner/tree/00c5b007f70094b5989ed219bc33ac2c55e01e41}{\texttt{00c5b00}}}.

For extracting the variability from the \textit{C source code} we chose Pilztaker \cite{Pilztaker}, a tool based on Undertaker 1.6.1 \cite{Undertaker} (commit from 02.06.2016\footnote{commit hash \href{https://vamos.informatik.uni-erlangen.de/trac/undertaker/browser?rev=40fcc5d54c66886bd3239a63c0a5e61862079ed1}{\texttt{40fcc5d}}}). This tool extracts each conditional compilation directive of the C preprocessor, that is found inside the C source files of the Linux kernel. This hierarchical structure is then translated into presence conditions. However, Pilztaker can not handle header files, that are included inside of C source files. It also can not find any usage of variability that is not inside \ifDefBlock{s} (for example, in macro expansions). To bypass these limitations, we filter out each variability variable that is used in headers or outside of \#ifdef blocks; these variables are not considered in any further analysis. We opted for this tool in contrast to other available tools, since including a macro-aware parsing lead to too complex constraints for the manual analysis of the identified results. Also, its limitations can be easily detected and affected variables can be avoided, in order to not produce false positives later on.

The \textit{mismatch detector} \cite{CMDetector} takes these three inputs and searches for configuration mismatches. To do that, the file presence conditions are combined with the code presence conditions. A mapping of each Kconfig variable, together with the presence condition they are used in, is created. The feature effects of each variable are calculated from this mapping, and finally a SAT-solver is applied to detect possible configuration mismatches, as described at the beginning of this section. For each configuration mismatch found, the tool outputs the problematic variable and minimal partial configurations, under which the problem is reproducible.


\section{Analysis \secSize{3,5 - 4}}
\label{sec:analysis}
Our tooling reported 247 potential configuration mismatches for the x86 architecture of the Linux kernel version 4.4.1. Due to the SAT-based analysis, \texttt{tristate} variables were treated as 2 independent variables. This means that our tooling reported two distinct results, if it detected a configuration mismatch for the selection as a loadable module (\texttt{m}) and for the permanent selection (\texttt{y}). If only one of these selections lead to a mismatch or if the variable was of type \texttt{bool}, the tooling reported only one result.

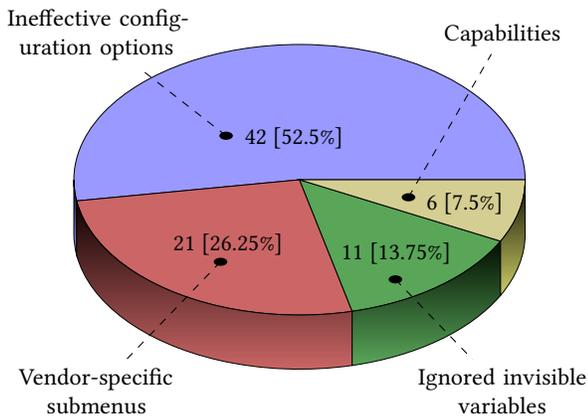
\begin{figure}[!tb]
  \begin{tikzpicture}[scale=3]
    \begin{scope}[xscale=1,yscale=3/5]
      \slice{0}{525/1000 * 360}{\nIneffective{} [52.5\%]}{blue}{40}{white}
      \slice{525/1000 * 360}{7875/10000 * 360}{\nSubmenus{} [26.25\%]}{red}{50}{gray!80!white}
      \slice{7875/10000 * 360}{925/1000 * 360}{\nInvisibles{} [13.75\%]}{green}{30}{gray}
      \slice{925/1000 * 360}{360}{\nHWvariables{} [7.5\%]}{yellow}{50}{gray!80!white}
      
      \pieLabel{135:0.46cm}{130:1.4cm}{Ineffective configuration options}{text width=2.75cm,align=center}
      \pieLabel{240:0.7cm}{240:1.8cm}{Vendor-specific submenus}{text width=2.25cm,align=center}
      \pieLabel{300:0.85cm}{300:1.8cm}{Ignored invisible variables}{text width=2.25cm,align=center}
      \pieLabel{345:0.5cm}{50:1.4cm}{Capabilities}{text width=2.75cm,align=center}
    \end{scope}
  \end{tikzpicture}
  \caption{Distribution of analysis results.}
  \label{fig:distribution}
\end{figure}

In the first step, we excluded 150 results for the manual analysis. This was necessary for the following reasons:
\begin{itemize}
	\item 89 of them refer to variables, which were also used in header files. We excluded these results to avoid false positive results as we could not reliably calculate feature effects for these variables without a full support of the C preprocessor syntax they were to numerous for manual post-processing.
  \item 48 findings were related to variables, which are dependent of architecture-specific variables of another architecture. These variables are also used in common code artifacts, e.g., in drivers. A sound analysis of configuration mismatches would require to include the variability models of other architectures in Formula~\ref{eq:sat test}.
  \item 3 variables controlled compiler settings during the build, which could not be expressed as presence conditions in our analysis.
  \item 10 identified configuration mismatches had very complex feature effect constraints involving more than \ToDoInline{12} variables. Due to the complexity of the constraints, our tool was unable to enumerate all required partial configurations, which is an NP-hard problem. However, our manual analysis relied on these partial configurations. For this reason we excluded these configurations mismatches from the manual analysis.
\end{itemize}
We applied a manual analysis for all of the remaining 97 findings:
\begin{itemize}
	\item \nSmells{} of these results could be verified as configuration mismatches and are categorized below. These results are related to \nVars{} distinct variables of the Kconfig model (\nBools{} variables of type \texttt{bool} and \nTristates{} variables of type \texttt{tristate}). 
  \item 15 of these results were caused by wrong intermediate results of KbuildMiner. These could be traced down to special cases of the make-files, which were not handled correctly.
  \item 1 was related to the usage of a presence condition in a C-file, which used a variable with a prefix \texttt{CONFIG\_}, but was not defined by the Kconfig model.
  \item 1 result was related to an incomplete handling of Kconfig \texttt{choice}s (comparable with feature alternatives). The tooling we used did not consider that the deselection of one variable in the choice always results in the selection of another one. In this specific result, all but one element in the choice always had a feature effect. Our tool reported the one without a feature effect as a mismatched configuration, not considering that its selection always had an indirect feature effect via the other choice items.
\end{itemize}
The manual analysis of the \nSmells{} configuration mismatches revealed 4 different categories, which we describe in detail in the following subsections. These are: Ineffective configuration options, vendor-specific submenus, ignored invisible variables, and capabilities.

The distribution is as follows (cf.\ Figure~\ref{fig:distribution}): \nIneffective{} of the results point to situations were the configuring user is able to configure Kconfig variables without any effect. \nSubmenus{} of the results are used to structure the variability model without affecting the final product at all. \nInvisibles{} results point to configuration issues with invisible Kconfig variables. The remaining \nHWvariables{} results are related to variables for modeling platform dependent capabilities.

Each subsection starts with a conceptual description of a specific type of configuration mismatch, followed by a meaningful example, which we took from the Linux kernel version 4.4.1. Based on the presented example, we discuss the criticality of the identified category.

  \subsection{Ineffective Configuration Options}
In this section, we discuss a type of configuration mismatches, which is a prime example of the induced problems. Configuration options are defined and need to be considered by the user. Contrary to their description, the selection of those variables does not affect the final product under certain conditions. As a consequence, the configuration process takes longer than needed and may actually lead to different results than those expected as the user may choose a certain option, while the final product does not even contain the functionality. This problem arises if visible variables of the variability model are not constrained in the same manner as the associated code artifacts. In \cite{El-SharkawyKrafczykSchmid16} we have already presented an example, which allows a detailed configuration of memory sticks during suspension, even if the user disabled support for suspension before. Below we show another meaningful example: the variable \ATHK{} in the wireless network drivers from Atheros.

\begin{figure}[tb]
	\centering
\begin{lstlisting}[language=kconfig,caption={Excerpt of ineffective configuration options (Kconfig).},label=lst:kconfig_smell_example]
config ATH5K
    tristate "Atheros 5xxx wireless cards support"
    depends on (PCI || ATH25) && MAC80211
    select ATH5K_PCI if !ATH25
    ...

config ATH5K_PCI
    bool "Atheros 5xxx PCI bus support"
    depends on (!ATH25 && PCI)
    ...
    
\end{lstlisting}
\end{figure}

In Listing~\ref{lst:kconfig_smell_example}, we show the relevant excerpt of the Kconfig model for Atheros wireless cards of the 5000 series. If the user selects the variable responsible for the inclusion of the complete driver (\ATH), it automatically forces the boolean variable \ATHK{} to be selected as \texttt{y}, if the \ATHTwentyFive{} is deselected (cf.\ Line~\ToDoInline{4}). The later is a variable of the ``MIPS'' architecture and, thus, is deselected in all other architectures. However, \ATHK{} has also a prompt (cf.\ Line~\ToDoInline{8}). This allows the manual configuration of this variable, if it was not selected before. 
As a consequence, the user can configure \ATHK{} only if \ATH{} is deselected.

\begin{figure}[!hbt]
	\centering
\begin{lstlisting}[language=kbuild,caption={Excerpt of ineffective configuration options (Kbuild).},label=lst:kbuild_smell_example]
#drivers/net/wireless/ath/ath5k/Makefile
ath5k-$(CONFIG_ATH5K_PCI)	+= pci.o
obj-$(CONFIG_ATH5K)		+= ath5k.o
...
\end{lstlisting}
  \vspace{-2ex}
\end{figure}

Listing~\ref{lst:kbuild_smell_example} shows the relevant part of Kbuild for the compilation of the related C-files. The only usage of \ATHK{} inside the solution space is in Line~\ToDoInline{2}. The variable is used to control the conditional compilation of \texttt{pci.c} as part of the \texttt{ath5k} object. The result of this compilation is statically linked into the kernel or compiled as loadable module depending on the value of \ATH{} (cf.\ Line~\ToDoInline{3}). 
As a consequence, \ATHK{} is only considered for compilation if \ATH{} is selected.

The combination of the modeled and the implemented dependencies result in a configuration mismatch. More precisely, Kconfig facilitates the configuration of \ATHK{} exactly if it has no impact. This is obviously a fault.

This case is obviously related to an incorrect usage of Kconfig, since the relationship between both variables is the same in the problem space and in the solution space (\ATHK{} is used for a detailed configuration of \ATH). However, without a complete knowledge about all existing \ATH{} cards, it remains unclear how to solve this issue. There exist two possible solutions:
\begin{itemize}
  \item The usage of a conditional \texttt{default} value would be more useful if not all but most of these cards are built for PCI (soft constraint), to facilitate overwriting the default selection.
  \item The usage of the \texttt{select} constraint is correct if all \ATH{} cards are PCI based (hard constraint). However in contrast to the current situation, \ATHK{} should be made invisible to avoid illegal configurations. This is also recommended by the Kconfig language specification, which discourages the selection of visible variables \cite{KConfig_Language}.
\end{itemize}

While this example is obviously related to a missing dependency in the Kconfig model, this is not necessarily the case for all instances of this category. In general, a configuration mismatch of this category may point to a missing constraint of the problem space or to an over-constrained solution space.

  \subsection{Vendor-Specific Submenus}
Some parts of the Kconfig model use additional variables to group product-specific variables of a vendor. These variables can be identified by their name as they usually contain the name of vendor instead of a specific product or functionality. Vendor-specific variables do not directly affect the product derivation. However, these variables must be selected to facilitate the selection of the product-specific variables. Thus, the purpose of these vendor-specific variables is to form a selectable submenu. As a consequence, for each of these vendor-specific variables exists a redundant configuration: the deselected vendor-specific variable and the selected vendor-specific variables without any selected product-specific variables form the same product. The description of the vendor-specific variables points out that this behavior is actually intended as they contain the hint: \emph{``Note that the answer to this question doesn't directly affect the kernel: saying N will just cause the configurator to skip all the questions about <VENDOR> cards. If you say Y, you will be asked for your specific card in the following questions.''} Examples exist for Google, Cisco, Qualcomm, and Samsung, but also for other vendors. Below we present a representative example based on the vendor-specific submenu for Ethernet devices of Qualcomm.

\begin{figure}[tb]
	\centering
\begin{lstlisting}[language=kconfig,caption={Excerpt of a vendors-specific Submenu (Kconfig).},label=lst:kconfig_categories_example]
config NET_VENDOR_QUALCOMM
    bool "Qualcomm devices"
    default y
    ---help---
     If you have a network (Ethernet) card belonging to this class, say Y.

    Note that the answer to this question doesn't directly affect the kernel: saying N will just cause the configurator to skip all the questions about Qualcomm cards. If you say Y, you will be asked for your specific card in the following questions.

if NET_VENDOR_QUALCOMM

config QCA7000
    tristate "Qualcomm Atheros QCA7000 support"
    depends on SPI_MASTER && OF
    ---help---
     This SPI protocol driver supports the Qualcomm Atheros QCA7000.

    To compile this driver as a module, choose M here. The module will be called qcaspi.

endif # NET_VENDOR_QUALCOMM
\end{lstlisting}
  \vspace{-2ex}
\end{figure}

Listing~\ref{lst:kconfig_categories_example} shows the information relevant information of the Kconfig file in \texttt{drivers/net/ethernet/qualcomm}. This file contains exactly two variables; a vendor specific-variable to group all Ethernet drivers from Qualcomm (\vendorQualcomm) and one grouped product-specific variable (\QCA). Through the surrounding if-Statement in lines \ToDoInline{9} and \ToDoInline{19}, \QCA{} will be nested below \vendorQualcomm{} and also only be selectable if the vendor-specific variable was selected before. Line \ToDoInline{3} defines a \texttt{default} value to make the vendor-specific submenu with its nested variable visible as long as the configuring user does not deselect it.

\begin{figure}[b]
	\centering
\begin{lstlisting}[language=kbuild,caption={Excerpt of a vendor-specific submenu (Kbuild).},label=lst:kbuild_categories_example]
#drivers/net/ethernet/Makefile
obj-$(CONFIG_NET_VENDOR_QUALCOMM) += qualcomm/

#drivers/net/ethernet/qualcomm/Makefile
obj-$(CONFIG_QCA7000) += qcaspi.o
qcaspi-objs := qca_spi.o qca_framing.o qca_7k.o qca_debug.o
...
\end{lstlisting}
  \vspace{-2ex}
\end{figure}

Listing~\ref{lst:kbuild_categories_example} shows the relevant structure of the involved build-files. At first glance it looks though \vendorQualcomm{} affects the derivation process, since it is used in Line~\ToDoInline{2}. However, this line includes Qualcomm's subfolder together with its build-file. This build-file specifies that all nested files are compiled only if \QCA{} has been selected and, thus, selecting \vendorQualcomm{} without also selecting \QCA{} becomes meaningless. Somebody could argue that the vendor-specific variables ensure that no elements of the subfolders may be included accidentally, in case there are elements marked for compilation without any condition.

\begin{figure}[tb]
	\centering
		\includegraphics[width=\columnwidth]{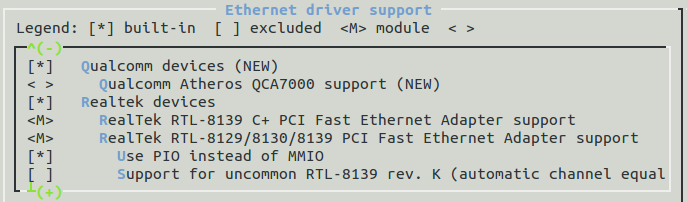}
  \caption{UI presentation of the vendor-specific submenu for \vendorQualcomm{} in Kconfig.}
  \label{fig:MenuOfcategories}
\end{figure}

Figure~\ref{fig:MenuOfcategories} shows how the vendor-specific submenu for \vendorQualcomm{} is rendered in Kconfig's configuration tool. By default, all vendor-specific submenus are visible together with their product-specific variables. The possibility of deselecting the vendor-specific variables introduces only a kind of ``folding'' mechanism.

According to our evaluation, these vendor-specific submenus seem not to be a problematic issue, even though they introduce redundant configurations. The description of the vendor-specific variables point out that this behavior is intended by the developer. Thus, these variables can be regarded as \textit{abstract features} \cite{ThumKastnerErgweg+11}.

  \subsection{Ignored Invisible Variables}
This category describes the situation where an invisible variable (without a prompt) is ignored during the instantiation process. These variables have a conditional default value or are selected by other variables and should not be changed by the configuring user. Thus, these variables are used to control features of the kernel, which are directly dependent on other features. Below we present a representative example based on Intel MID (Mobile Internet Device) platform systems.


Listing~\ref{lst:kconfig_deselect_example} shows the information relevant for the configuration of serial ports on Intel MID platforms. During the configuration of an Intel MID platform, the \DMAPCI{} should automatically be selected if the configuring user also selects the \SERIAL{} variable (by means of the \texttt{select} constraint in Line~\ToDoInline{5}). The \DMAPCI{} contains an additional \texttt{depends on} constraint to restrict its visibility and range, if the condition is not met. However in Kconfig, in conflicting situations \texttt{select} constraints have a higher priority than \texttt{depends on} constraints. Thus, \DMAPCI{} will automatically be selected even if \DMA{} is disabled.

\begin{figure}[tb]
	\centering
\begin{lstlisting}[language=kconfig,caption={Excerpt of ignored invisible variables (Kconfig).},label=lst:kconfig_deselect_example]
config SERIAL_8250_MID
    tristate "Support for serial ports on Intel MID platforms"
    depends on SERIAL_8250 && PCI
    select HSU_DMA      if SERIAL_8250_DMA
    select HSU_DMA_PCI  if X86_INTEL_MID
    ...

config HSU_DMA
	tristate
	select DMA_ENGINE
	select DMA_VIRTUAL_CHANNELS

config HSU_DMA_PCI
	tristate
	depends on HSU_DMA && PCI
\end{lstlisting}
  \vspace{-2.5ex}
\end{figure}

\begin{figure}[b]
  \centering
\begin{lstlisting}[language=kbuild,caption={Excerpt of ignored invisible variables (Kbuild).},label=lst:kbuild_deselect_example]
#drivers/dma/Makefile
obj-$(CONFIG_HSU_DMA) += hsu/

#drivers/dma/hsu/Makefile
obj-$(CONFIG_HSU_DMA_PCI)	+= hsu_dma_pci.o
...
\end{lstlisting}
\end{figure}

Listing~\ref{lst:kbuild_deselect_example} shows how the involved build files force the compliance of the (ignored) \texttt{depends on} constraint. The compilation of the relevant files depends on the selection of the \DMAPCI{} variable, but they are also stored in the sub folder \texttt{hsu}. The consideration of the complete folder further dependents of the \DMA{} variable. As a consequence, the selection of the \DMAPCI{} variable becomes irrelevant if \DMA{} is not also selected.

It is questionable if this behavior is really intended by the developers. The names of the involved variables together with the ignored \texttt{depends on} constraint indicate that the implemented behavior was intended for this example. But then the question remains open why the constraint was modeled this way. The situation could easily be fixed, if the condition of the \texttt{depends on} constraint would be added to the condition of the \texttt{select} constraint in Line~\ToDoInline{5}.

This category indicates that some kernel developer are not aware of the full semantics of the Kconfig language. The reason for this could be the weak documentation or the confusing behavior of the precedence of different constraint types. Instead the developer use the build mechanisms to enforce the correct product derivation.

\vspace{-0.5ex}
\subsection{Capabilities}
This section is similar to the previous one, but differs insofar as the identified results use variables which are used to describe platform-dependent capabilities, e.g., supported features of a CPU architecture. The names of these variables usually are in the form of ``\texttt{<plat\-form>\_HAVE/HAS/SUPPORTS/WANTS/\ldots\_<function>}''. Below, we present an example of these capabilities based on the Xen hypervisor.



Listing~\ref{lst:kconfig_have_example} presents an excerpt of the Kconfig model related to the configuration of Xen. Lines~1--7 are specific for the x86 architecture, the lines below are from the common part. The architecture-specific part exists in a similar way also for the other architectures. For instance, the ARM architecture defines the configurable variable \XEN{} 	analogous, but does not specify the \texttt{select} constraint of Line~5. The selected configuration option (cf.\ Line~15) is such an invisible variable to model the provided capabilities of the Xen platform: It specifies that the x86-specific implementation of Xen provides a virtual performance monitoring unit. Independent from the architecture-specific settings, the user is able to generally configure monitoring via the \XENSYS{} variable (cf.\ Line~10).

In Kbuild, the variable \XENSYS{} is used to select the C-file for compilation, which realizes the monitoring feature. This part is architecture independent. Inside the C-file, \XENVPMU{} is used in preprocessor statements to augment the logging output, if a virtual performance monitoring unit is available in the current platform.

In this category the configuration mismatch appears because additional variables are introduced to represent capabilities. This appears to be an explicit design decision by the people constructing the variability model, hence, we do not regard it as problematic. Rather, this can be seen as an evolution provision. Indeed, such a configuration mismatch need not be present in other architectures. 

\begin{figure}[tb]
	\centering
\begin{lstlisting}[language=kconfig,caption={Excerpts of Xen-specific Kconfig files (Lines~\ToDoInline{1}--\ToDoInline{7} are specific for the x86 architecture).},label=lst:kconfig_have_example]
# arch/x86/xen/Kconfig
config XEN
    bool "Xen guest support"
    ...
    select XEN_HAVE_VPMU
    help
     This is the Linux Xen port. Enabling this will allow the kernel to boot [...] under the Xen hypervisor.

# drivers/xen/Kconfig
config XEN_SYS_HYPERVISOR
    bool "Create xen entries under /sys/hypervisor"
    depends on SYSFS

...
config XEN_HAVE_VPMU
    bool
\end{lstlisting}
\end{figure}

\FPeval{\nInvisNoprob}{\nHWvariables}
\FPeval{\nInvisProblem}{\nInvisibles}
\FPeval{\nVisNoprob}{clip(\nSubmenus)}
\FPeval{\nVisProblem}{clip(\nIneffective)}
\FPeval{\nTotalNoprob}{clip(\nVisNoprob + \nInvisNoprob)}
\FPeval{\nTotalProb}{clip(\nVisProblem + \nInvisProblem)}

\vspace{-1ex}
\section{Discussion \secSize{3/4}}
\label{sec:Discussion}
In this section we discuss our findings in the relation to  our research questions.  Figure~\ref{fig:smell categorization} provides a summary.

\vspace{-1ex}
\subsection{\RQLayerOne{1}: Severity of Findings}
\label{sec:rq_severity}
According to our analysis  \nTotalProb{} of \nSmells{} (66.25\%) analyzed configuration mismatches are real problems in the sense that the final product will have a different functionality than would be expected according to the configuration. This is depicted in  Figure~\ref{fig:smell categorization}.
 Without a deep understanding of the domain knowledge it remains unclear whether these problems are related to an over-constrained solution space or an under-constrained problem space.

The identified problems can be further subdivided into two different categories according to their impact on  product derivation:
\begin{itemize}
	\item 79.25\% of the problematic configuration mismatches are related to visible variables (\textit{ineffective configuration options}). 
	The user must configure configurations options, 
	 which have absolutely no impact  on the configuration result under certain conditions. Thus, the configuration process takes longer than needed and leads to different results than those expected as the user may choose a certain option, while the final product does not  contain the functionality. 
  \item 20.75\% are \textit{ignored invisible variables}. These variables should be used to realize technical settings depending on the current configuration. Instead of   proper usage of constraints to ensure a correct selection of those variables, the developers rely on implementation techniques to ensure correct functionality of the derived kernel. As a result, these variables become useless.
\end{itemize}

\newcommand{\layerDistance}{0.275}
\FPeval{\levelDistance}{clip(2 * \layerDistance)}

\begin{figure}[tb]
  \begin{tikzpicture}[
      smell/.style={rectangle,thick,draw,top color=BlueDown!5,bottom color=blue!35,text width=2cm,text centered,font=\sffamily,anchor=north},
      category/.style={smell, rounded corners},
      problem/.style={smell, top color=red!5, bottom color=red!30}, 
      fine/.style={smell, top color=green!10, bottom color=green!20},
      edge from parent/.style={very thick,draw=black!70},
      edge from parent path={(\tikzparentnode.south) -- ++(0,-\layerDistance cm) -| (\tikzchildnode.north)},
      level 1/.style={sibling distance=4cm,level distance=\levelDistance cm, growth parent anchor=south,nodes=smell},
      level 2/.style={sibling distance=2.25cm},
      level 3/.style={sibling distance=2.05cm,level distance=0.6cm}
      ]
      
      \node (s1) [category, text width=3.5cm] {Configuration Mismatch\\\nSmells}
        child {\categoryDesc{v1}{category, text width=2.3cm, xshift=2mm}{Visible Variables}{63}{24}{39}
          child {\smellDesc{v2}{problem, text width=2.8cm, xshift=-4mm}{Ineffective C.\ Opt.}{42}{23}{19}
          }
          child {\smellDesc{v5}{fine, text width=1.5cm, xshift=-2mm}{Submenus}{21}{1}{20}}
        }   
        child {\categoryDesc{i1}{category, text width=2.5cm}{Invisible Variables}{17}{8}{9}
          child {\smellDesc{i2}{problem, text width=1.4cm}{Ignored}{11}{2}{9}}
          child {\smellDesc{i3}{fine, text width=1.5cm, xshift=-5mm}{Capabilities}{6}{6}{0}}
        }
      ;
      
      \node (l0) [anchor=north, xshift=-3.95cm, very thin, rectangle, text width=1.6cm, draw] {\footnotesize Legend:\\[.54cm]$ $};
      \node (l1) [problem, text width=1.45cm, xshift=-3.95cm, yshift=-.37cm] {\footnotesize Problematic};
      \node (l2) [fine, text width=1.45cm, xshift=-3.95cm, yshift=-.81cm] {\footnotesize Unproblematic};
      
  \end{tikzpicture}
  \caption{Categorization of our findings (C:\ No.\ of findings used in code files; Kb:\ No.\ of findings used in Kbuild).}
  \label{fig:smell categorization}
  \vspace{1em}
\end{figure}
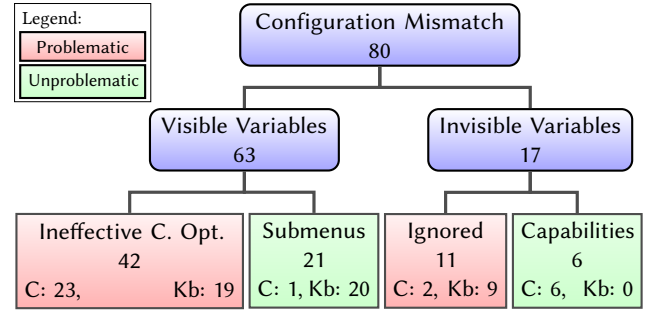

%

\subsection{\RQLayerOne{2}: Technical Characteristics of Variables}
Here, we discuss the characteristics of the variables causing the configuration mismatches. We observed the following distribution (cf.\ Table~\ref{tab:Technical differentiation of variables}):

\begin{description}
	\item[Type:] In Kconfig, all variables are typed. While the used tools are capable of handling  Strings and numerical variables also in propositional logic \cite{El-SharkawyKrafczykSchmid15}, we did not identify any configuration mismatches triggered by such variables. This is probably related to the low occurrence of those variables. A study revealed that only 6\% of the variables in the variability model of Linux are of those types, while 36\% are of type \texttt{bool} and 58\% are of type \texttt{tristate} \cite{BergerSheLotufo+13}.
  
  The 80 results of our analysis are related to 68 variables. \nBools{} variables are of type \texttt{bool} (79.4\%) and \nTristates{} variables are of type \texttt{tristate} (20.6\%). Most of the reported \texttt{tristate} variables had a configuration mismatch for both value assignments \texttt{m} and \texttt{y}. For two of these \texttt{tristate} variables, only the selection as a loadable module (\texttt{m}) was problematic.
  \item[Visibility:] Another characteristic is that not all defined variables are visible to the end-user (i.e., variables that do not have prompt). These variables can only be configured through their dependencies. Therefore, correct constraint modeling becomes even more important for those variables to be effective. This differentiation is presented in the second layer of Figure~\ref{fig:smell categorization} as it serves as a basis of our classification of configuration mismatches.\ks{So what? Here fehlt die Konsequenz? auf die werte wird auch nicht eingegangen.}
\end{description}

\begin{table}[tbh]
	\vspace{1ex}
  \centering
		\begin{tabular}{|l|r|r|l|r|r|}
      \hline
      \textbf{Type}&\textbf{No.}&\textbf{Dist.}&\textbf{Visibility}&\textbf{No.}&\textbf{Dist.}\\
      \hline
      \multirow{2}{*}{\textbf{Boolean}}&\multirow{2}{*}{54}&\multirow{2}{*}{79.4\%}&Visible&44&64.71\%\\
      \cline{4-6}
      &&&Invisible&10&14.71\%\\
      \hline
      \multirow{2}{*}{\textbf{Tristate}}&\multirow{2}{*}{14}&\multirow{2}{*}{20.6\%}&Visible&12&17.65\%\\
      \cline{4-6}
      &&&Invisible&2&2.94\%\\
      \hline
      \textbf{Total}&\textbf{68}&\textbf{100\%}&\textbf{Total}&\textbf{68}&\textbf{100\%}\\
      \hline
		\end{tabular}
    \caption{Technical differentiation of involved variables and their distribution.}
    \label{tab:Technical differentiation of variables}
    \vspace{-2ex}
\end{table}

We can see that Boolean variables are significantly more involved in mismatches than \texttt{tristate} variables. Other types do not occur as mismatches. Also more variables with a prompt lead to configuration mismatches than invisible variables, which is the reason for the high ratio on \textit{ineffective configuration options} (cf.\ Section~\ref{sec:rq_severity}).


\subsection{\RQLayerOne{3}: Conceptual Differentiation}
In Section~\ref{sec:analysis}, we identified 4 categories of configuration mismatches. These categories form the third layer of the tree in Figure~\ref{fig:smell categorization}. Independently of their technical characterization, these 4 categories differ in their impact with respect to the product derivation:
\begin{itemize}
	\item 27 results are categorized as unproblematic. 21 of them are used to define submenus in Kconfig and 6 are used to model capabilities, which are not needed for certain configurations.
  \item 53 results point to a problematic divergence between modeled and implemented dependencies, which were already discussed in Section~\ref{sec:rq_severity}. These two categories affect different roles during the development: Ineffective configuration options are visible variables, which need to be considered by the configuring user without any impact on the product derivation, while ignored invisible variables affect developers only.
\end{itemize}

Even if some configuration mismatches seem to be desired by the developer, we see that most of the mismatches are critical and bear the risk of a wrong product derivation.

\subsection{\ref{rq:spaces}: Involved Spaces}

Our analysis revealed that all analyzed configuration mismatches are related to variables, which are either used in code artifacts or in Kbuild, but not in both.
This is surprising as approximately 19\% of all Kconfig variables are used to configure code and Kbuild artifacts together \cite{CMDetector}. However, our result may be not fully comparable as we excluded all variables from our analysis, which are also used in header files. Table~\ref{tab:Number of involved spaces} presents the absolute and relative distribution of our findings related to code and build files. It can be seen that the distribution for problematic variables is not the same as for unproblematic ones. We do not have a strong explanation for this divergence, hence, we just report it here. 
\vspace{-2ex}

\begin{table}[htb]
	\centering
		\begin{tabular}{|l|r|r|r|r|r|r|}
      \hline
      &\multicolumn{2}{c|}{\textbf{Code}}&\multicolumn{2}{c|}{\textbf{Kbuild}}&\multicolumn{2}{c|}{\textbf{Total}}\\
      &\textbf{No.}&\textbf{Dist.}&\textbf{No.}&\textbf{Dist.}&\textbf{No.}&\textbf{Dist.}\\
      \hline
      \textbf{Problematic}&25&47.17\%&28&52.83\%&53&100\%\\
      \hline
      \textbf{Unproblematic}&7&25.93\%&20&74.07\%&27&100\%\\
      \hline
      \textbf{Total}&\textbf{32}&\textbf{40\%}&\textbf{48}&\textbf{60\%}&\textbf{80}&\textbf{100\%}\\
      \hline
		\end{tabular}
    \caption{Results related to solution space.}
    \label{tab:Number of involved spaces}
\end{table}

\vspace{-7ex}
\section{Threats to Validity \secSize{1/2 - 3/4}}
\label{sec:threats}
In this section, we discuss threats to validity regarding our analysis, which consists of an \textit{automatic identification of configuration mismatches} and a \textit{manual assessment} of the results. Beyond this discussion, we made the tool for the detection of configuration mismatches and the raw data of our analysis publicly available \cite{CMDetector} to facilitate the reproduction of our analysis and to allow the independent verification of our findings.

\textbf{Internal validity.} The automatic analysis of Linux is not a trivial task. Linux consists of complex artifacts written in several different, decades-old languages \cite{NadiBergerKastner+15}. The variability modeling approach consists of many undocumented corner cases, which hamper the automatic analysis \cite{El-SharkawyKrafczykSchmid15}. To overcome these issues, we built our toolchain for the \textit{automatic identification of configuration mismatches} on existing research tools, which have already shown that they perform very well. Then, we applied a manual analysis to all results to exclude false positive reports from further analysis. However, the total number of unidentified configuration mismatches remains unclear as we also had to exclude some variables from our consideration (see construct validity)

All results, which have been confirmed as real configuration mismatches, have been \textit{manually assessed} by one author. Some samples were verified by a second author to ensure that the results are reproducible and correct. We discussed each of the identified categories to rate its general criticality. However, we omit a further analysis whether the results are related to an over-constrained solution space or to missing constraints of the problem space, because this would require extremely detailed domain knowledge of the implementation of Linux.


\textbf{External validity.} The approach presented in this paper was explicitly designed to analyze Kconfig-based software product lines with C-code. Most of the used tools have already been applied to other Kconfig-based product lines, like BusyBox or uClibc. However, the analysis of these systems require additional effort to adapt the underlying tooling, because the systems differ in how they realize Kconfig. For instance, Busybox stores artifact-specific variability in the related code artifacts instead of separate configuration files as it is done in Linux \cite{Busybox}.

We chose Linux as a case study, because it was already often the object of investigation in scientific research. Further, the developers spent much effort to build a huge infrastructure to test each new change. It monitors several development trees and also the mailing list. The test infrastructure tests changes on more than 100 different kernel configurations and applies statical tests \cite{0Day-Tests}. For this reason, we expect other systems to contain more configuration issues relative to their size and complexity.
Our expectation is underpinned by the analysis of Nadi et al.~\cite{NadiBergerKastner+15}, which revealed that in Linux most of the identified feature effects are also modeled as constraints in the variability model. Only 4\% of the identified feature effect conditions are not covered by the variability model and, thus, are configuration mismatches. The relative number of identified configuration mismatches was much higher for the other case studies (21\% -- 31\%).



\textbf{Construct validity.} We used the technique of feature effects \cite{NadiBergerKastner+15} to gather constraints from the variability model. This technique requires the knowledge of all presence conditions a variable is used in to reliably calculate its dependencies. Missing presence conditions may result in the calculation of a too restrictive constraint and, thus, in a false positive report. Thus, we omitted all variables from further analysis, for which we could not compute reliable feature effect constraints. These are variables used in header-files or which are involved in too complex presence conditions. Even if we excluded such variables in the first step from our analysis, there exist still enough configuration mismatches to obtain meaningful results.


\textbf{Conclusion validity.} The classification of configuration mismatches is a subjective process, which we did based on the \textit{manual assessment}. The categorization was discussed by all three authors of the paper to make the classification as objective as possible. Independent of this categorization, we presented two conceptual different examples, which demonstrate that configuration mismatches may point to potential kernel misconfigurations.

Answers to the research questions were done based on our experience with Linux and the observations we did. However, the analysis of the involved spaces is mitigated by the fact that we excluded variables from our analysis, which are also used in header files. Considering these variables may change the relative distribution of the findings in Table~\ref{tab:Number of involved spaces}.



\vspace{-1ex}
\section{Conclusion}
\label{sec:Conclusion}
In this paper, we identified configuration mismatches in Linux and applied a manual analysis to improve the understanding of their causes. We have shown that the bulk of these configuration mismatches point to real problems \RQRef{rq:severity}, which may be the result of missing or defect constraints in the variability model or wrong dependencies in code or build artifacts \RQRef{rq:spaces}. The identified configuration mismatches can be differentiated technically based on the visibility and the type of the triggering variable \RQRefBrackets{\ref{rq:technical}}. The visibility also indicates the main stakeholders, which are affected by the configuration mismatch: Visible configuration mismatches concern the configuring users as they make the configuration process unnecessary complex and may end up in the wrong product. Invisible configuration mismatches result in superfluous parts of the variability model and the developers need to encode the dependencies in the solution space. This paper supports a better understanding of the criticality of configuration mismatches in general and informs the development of automated analysis tools to identify them \RQRef{rq:characteristics}.

The analysis of other Kconfig-based product lines and a macro-aware parsing remain future work. The analysis of Nadi et al.\ \cite{NadiBergerKastner+15} has shown that other systems contain much more configuration mismatches than Linux. These systems have not been analyzed as thoroughly as Linux. For this reason, we expect more insights from these systems. In particular we expect deeper insights about their severity. A macro-aware processing makes the manual analysis more complex, which is needed to classify the impact of the configuration mismatch. However, this would also enable the inclusion of variables, which are used in header files and, thus, would make the analysis more comprehensive.
\begin{acks}
  This work is partially supported by the Evoline project, funded by the \grantsponsor{SPP1593}{DFG (German Research Foundation)}{http://www.dfg-spp1593.de/} under the Priority Programme \grantnum{SPP1593}{SPP 1593: Design For Future --- Managed Software Evolution} and by the ITEA3 project $\text{REVaMP}^2$, funded by the \grantsponsor{01IS16042H}{BMBF (German Ministry of Research and Education)}{https://www.bmbf.de/} under grant \grantnum{01IS16042H}{01IS16042H}. Any opinions expressed herein are solely by the authors and not of the DFG or the BMBF.

\end{acks}

\bibliographystyle{ACM-Reference-Format}
\bibliography{literature} 

\end{document}